\documentclass[11pt, a4paper]{article}

\usepackage{jinstpub}
\usepackage[utf8]{inputenc}
\usepackage{geometry}
\usepackage{cleveref}
\usepackage{amsmath,amsthm,bm}
\usepackage{physics}
\usepackage{siunitx}
\usepackage{hyperref}
\usepackage{pdfpages}
\usepackage{url, doi}
\usepackage{braket}
\usepackage[version=3]{mhchem}
\usepackage{comment}
\usepackage{subcaption}
\title{An indirect geometry crystal time-of-flight spectrometer for FRM II}

\author[a]{R.~Tang,}
\author[a]{C.~Herb,}
\author[b, 1]{J.~Voigt%
\note{Corresponding Author.}}
\author[a, 2]{R.~Georgii%
\note{Corresponding Author.}}

\affiliation[a]{Heinz Maier-Leibnitz Zentrum, Technische Universität München\\
Lichtenbergstr.1, Garching, Germany}
\affiliation[b]{Jülich Center for Neutron Science, Forschungszentrum Jülich\\
Wilhelm-Johnen-Straße, 52428 Jülich, Germany}

\emailAdd{Robert.Georgii@frm2.tum.de}
\emailAdd{j.voigt@fz-juelich.de}

\abstract{We present a concept for an indirect geometry crystal time-of-flight spectrometer, which we propose for a source similar to the FRM-II reactor in Garching. Recently, crystal analyzer spectrometers at modern spallation sources have been proposed and are under construction. The secondary spectrometers of these instruments are evolutions of the flat cone multi-analyzer for three-axis spectrometers (TAS). The instruments will provide exceptional reciprocal space coverage and intensity to map out the excitation landscape in novel materials. We will discuss the benefits of combining a time-of-flight primary spectrometer with a large crystal analyzer spectrometer at a continuous neutron source.  The dynamical range can be very flexibly matched to the requirements of the experiment without sacrificing the neutron intensity. At the same time, the chopper system allows a quasi-continuous variation of the initial energy resolution. The neutron delivery system of the proposed instrument is based on the novel nested mirror optics, which images neutrons from the position of the pulse cutting chopper representing a bright virtual source onto the sample. The spot size of less than 1 cm x 1 cm at the virtual source allows the realization of very short neutron pulses by the choppers, while the small and well-defined spot size at the sample position provides an excellent energy resolution of the secondary spectrometer.}

\arxivnumber{2405.09159}
\begin{document}

\maketitle
\flushbottom

\section{Science case}
Magnetic systems are fertile ground for the design of novel quantum and topologically non-trivial states characterized by exotic excitations. Recent examples include spin chain \cite{Tennant1995} and square-lattice low-dimensional antiferromagnets \cite{Piazza2015}, quantum spin liquid candidates \cite{Banerjee2017, Li2017}, spin-ice compounds \cite{Bramwell2001}, and unusual spin textures \cite{Muehl2009, Weber2022}. These systems are not only of fundamental interest but may also pave the way to new technologies. For example, skyrmion spin textures open new possibilities for data storage in race track memories \cite{Shu2023} and allow for the design of electronic--skyrmionic devices \cite{Zang2015}.

Key features of the ground state and finite-temperature behaviour of a magnetic system are captured by the spectrum of its excitations. The aforementioned systems reveal exotic excitations dissimilar to standard magnons that form narrow bands in conventional ferro- and antiferromagnets. Detecting exotic excitations is far more challenging, as they show broad distribution in the energy and momentum space. Typical examples are helimagnons, the excitations of helical magnets like in MnSi \cite{Weber2022}. They are observed at very low energies and extend over a wide range of momenta. The spin-ice and spin-liquid candidates among the 4f oxides entail weak magnetic interactions, thus showing spectral features at typical energy transfers of 1-2 meV. These features are intrinsically very broad, as they often reflect fractionalized excitations carrying spin-1/2 in contrast to spin-1 for conventional magnons. These latest trends are often combined with the fact that the systems are magnetically very dilute, and the samples are getting smaller and smaller in physical size. 
Such complex dispersion relations require long measurement times and great effort when measured on triple-axis spectrometers (TAS), which only access a single point in ($Q$, $E$)-space at a time. 
Furthermore, their dispersions occupy large volumes in the ($Q$, $E$)-space, making it time-consuming to explore them point by point using a conventional three-axis spectrometer.

Studying samples down to some mm$^3$ will allow us to study materials synthesized under conditions that will never produce large crystals, such as high-pressure synthesis and hydrothermal synthesis. This will lead to input from inelastic neutron scattering immediately after materials are discovered or directly lead to the discovery of materials. 

In soft matter, incoherent processes in molecules include vibrations, rotations, and diffusion. However, molecular dynamic computer models have evolved to describe molecule behaviour over the complete energy and wave vector range of inelastic neutron scattering. The instrument proposed here will allow for studies of collective (hence wave-vector-dependent) dynamics in soft matter. For example, in membranes, collective dynamics are believed to drive the transport of molecules, pore opening, membrane fusions, and protein-protein interactions \cite{Rhein2012}. Limiting the illumination to small sample volumes will allow for the separation of background from sample environments, and at a later stage, polarisation will allow the separation of coherent and incoherent motions. Furthermore, time-resolved studies of soft matter stimulated out of equilibrium using pump-probe techniques would become possible.
\section{Design}
Recently, secondary spectrometers originally developed for TAS instruments have been proposed as back-ends for inverted geometry time-of-flight spectrometers at pulsed neutron sources, namely MUSHROOM at ISIS \cite{BEWLEY2021165077} and BIFROST at the ESS \cite{ANDERSEN2020163402}.
They offer an even wider coverage of ($Q$, $E$)-space through the continuous initial neutron spectrum with the bandwidth
\begin{equation}
    \label{eq:Bandwidth_general}
    \Delta \lambda = \frac{h}{m_N}\frac{1}{L f}
\end{equation}
with the distance $L$ between the neutron source and the sample and the frequency $f$ of the source. The initial energy resolution is defined either by the pulse length of the neutron source or by neutron choppers.

Bewley \cite{BEWLEY2021165077} proposed an analyser composed of pyrolitic graphite (PG) crystals covering a wide range of the solid angle as seen from the sample.
The analyser crystals are arranged in a configuration reminiscent of a mushroom cap, facilitating the imaging of neutrons from the sample onto a focal ring, below which a position-sensitive detector bank detects them. This achieves a large acceptance of the analyzer crystal array, using crystals with a wide mosaic spread and recovering a good energy resolution via the above-described imaging process.  
For this design, the resolution depends on the size of a small illuminated sample area.

\begin{figure}[htp]
    \centering
    \includegraphics[width=\textwidth]{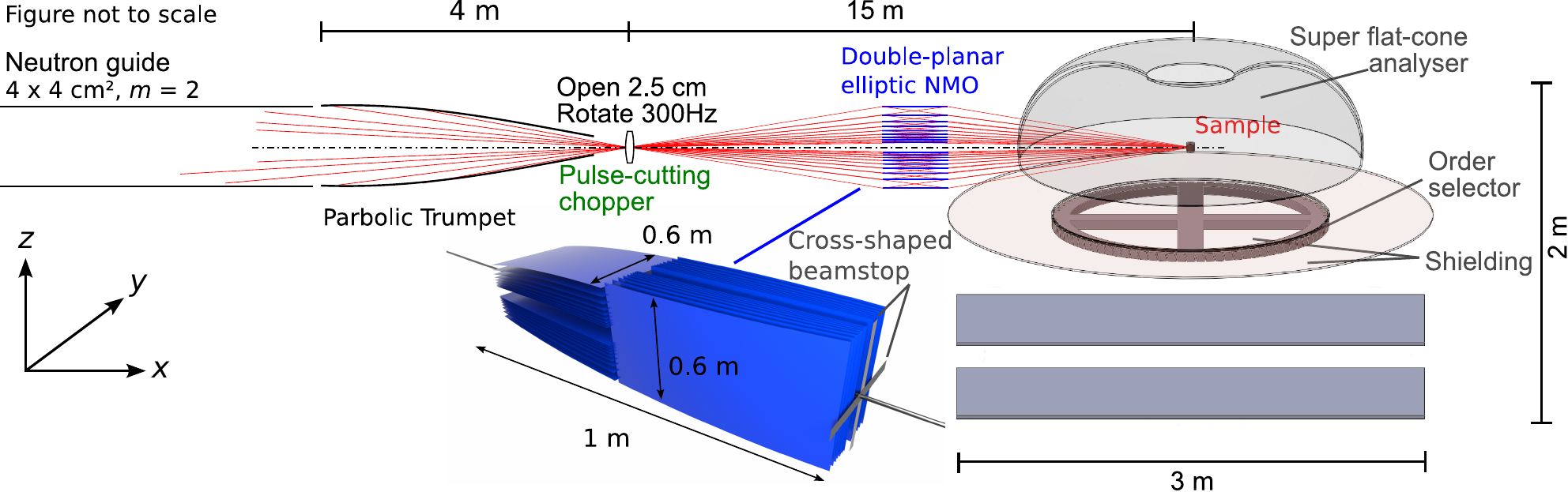}
    \caption{A not-to-scale overview of the instrument design. The neutrons are extracted from the reactor source by a neutron guide with an area of $40 \times \SI{40}{\milli\meter^2}$ which connects to a long (\SI{3.75}{\meter}) parabolic trumpet that reduces the extent of the beam to $10 \times \SI{10}{\milli\meter}$. At the nominal focal point of the parabola (located \SI{4}{\meter} downstream of its entrance), a pulse cutting and shaping chopper system with openings of $10\times\SI{10}{\milli\meter}$ is located, defining the dynamical range and the energy resolution of the primary spectrometer. An elliptic NMO transports the neutrons emitted from the pulse-cutting chopper over a distance of $2f = \SI{15}{\meter}$ to the sample while preserving the phase space at the chopper. Upstream of the NMO, a cross-shaped beamstop restricts the direct line of sight between the virtual source and the sample. Above the sample, the dome consisting of a large number of PG crystals is situated. This secondary spectrometer refocuses the neutrons emitted by the sample onto an order selector below which they are detected at the large position sensitive 2D detector (PSD). The PG dome is separated from the shielded detector array with a neutron absorber (light orange) to minimize spurious artefacts.}
    \label{Fig:Mushroom_Optics_Together}
\end{figure}

We propose a Mushroom-type spectrometer on the exit of one of the neutron guides at the continuous reactor source such as the MLZ (schematics see Fig. \ref{Fig:Mushroom_Optics_Together}). It will employ nested mirror optics to achieve a well-defined brilliant beamspot with an area not exceeding 10 mm $\times$ 10 mm, ideally suited for investigating novel single crystalline materials in complex sample environments.
In comparison to designs for pulsed sources with fixed frequency and hence bandwidth, the design allows for the independent tailoring of resolution and bandwidth through an advanced chopper system.

With a highly flexible primary chopper, our proposed spectrometer of Mushroom type would provide all necessary information within a much shorter time frame, only at the cost of a slightly reduced energy and momentum resolution as compared to a TAS and significantly more flexible in the choice of intensity and energy resolution than spectrometers at spallation sources. Furthermore, it can be built relatively compact as the main dimension of the secondary spectrometer is given by the dome of PG crystals above the sample. For the magnetic systems mentioned in the science case a possibility to reach low temperature and high magnetic fields is indispensable. Therefore, the PG dome has a hole of 50 cm in diameter at the top allowing for the employment of a typical closed cycle cryostat of the FRM 2 sample environment. Using a closed $^3$He insert, temperatures down to mK are achievable. Furthermore, a compact high-temperature superconducting magnet with a large opening angle can also be mounted there. It would also be possible to use high-pressure cells with a wider angular opening as currently being developed at FRM 2. Also, for the last two systems, the coverage of the ($Q$, $E$)-space is reduced due to the absorption of the material at larger takeoff angles.

The design can be augmented by a polarisation option at a later stage. In this case, the nested mirror optics will consist of polarising supermirrors, and a guide field will maintain the polarization of the neutrons upstream of the sample position. For bio and soft matter samples, this would already allow for the separation of coherent and incoherent intensity \cite{GASPAR201076}. Using an SEOP $^3$He polarising cell \cite{PhysRevLett.91.123003} with a sphere geometry after the sample would allow for spin analysing and separating magnetic and non-magnetic scattering and determining the chiral fraction. 

 Fig. \ref{Fig:Mushroom_Optics_Together} depicts the design of the (unpolarised) instrument. In the following, we first explore the design of the primary spectrometer, followed by describing the secondary crystal dome spectrometer. Finally, the overall resolution of the combined instrument is discussed, and a conclusion is given.

It is also worth mentioning that the current instrument suite of MLZ lacks a cold neutron instrument that provides full coverage of the out-of-plane ($Q, E$)- space.
\section{Neutron transport}
\label{sec:neut_trans}
The secondary spectrometer presented in \cref{sec:secondary_spectrometer}, relies on a compact illuminated sample area, with the scattered signal being analysed by the super flat-cone analyser.
Different sections in the analyzer scatter to different positions on the position-sensitive detection system determining the final wave vector with a resolution depending on the spot size of the beam at the sample and the spatial resolution of the detector.
Rather arbitrarily, we limit the sample area to $10\,\times$\SI{10}{\milli\meter^2}, matching the typical cross sections of single crystals, serving as the starting point for all subsequent considerations. 
While sensitive to the extent of the illuminated area, utilizing the design of the secondary spectrometer to first order decouples the instrument's energy resolution from the initial beam's divergence \cite{doi:10.1063/1.4901160}.

This observation invites the consideration of a modern neutron transport system tailored specifically towards small beams and large divergences. Rather recently, several of these concepts have been proposed and implemented. 
One approach is based on the Selene-type optics, which utilizes two subsequent elliptic mirrors to minimize the geometric aberrations. This design, employed at the AMOR reflectometer at SINQ \cite{Stahn201644} and foreseen for the ESS reflectometer ESTIA \cite{ISI:000290355100003, ANDERSEN2020163402}, enables the imaging of small beams
 extending $\leq \SI{1}{\milli\meter}$ in one direction with good brilliance transfer.
 
Additionally, we have explored the potential of elliptic nested mirror optics (NMO), which consist of laterally nested short elliptic reflective surfaces that provide good brilliance transfer and preserve the neutron phase space between the two focal points \cite{herbchristoph2022}. Furthermore, NMOs allow for the customization of the size and the divergence of the beam at the sample position by placing apertures close to the chopper system and the mirrors, respectively.
Considering the above, we opted for an NMO to transport neutrons from the pulse-cutting chopper, which serves as a virtual source, to the sample, as illustrated in Fig.~\ref{Fig:Mushroom_Optics_Together}. 
\par

Considering the large acceptance of the secondary spectrometer, we aim for a maximum divergence of $\SI{2}{^\circ}\times \SI{2}{^\circ}$ at the sample position\footnote{More precisely, the distribution of the horizontal and vertical divergence of the beam is desired to be uniform in the range $\SI{-2}{^\circ}\leq \alpha_\mathrm{h,\, v} \leq \SI{2}{^\circ}$.}. Due to the timing requirements detailed in section \ref{sec:choppersystem}, we selected a distance of \SI{15}{\meter} between the virtual source and the sample. This determines the symmetric focal lengths measured from the centre of the assembly coinciding with the semi-minor axis of the ellipses, $f= \SI{7.5}{\meter}$. By choosing the maximum divergence and the distance between the common focal points of the elliptic reflective surfaces comprising the NMO, the geometry of the NMO transport system is fixed.\par
To redirect the neutron beam in two dimensions, i.e., vertical and horizontal transport, we require two orthogonally oriented NMOs, which are placed behind each other on the optical axis. This device is referred to as a double-planar elliptic NMO (compare the inset of \cref{Fig:Mushroom_Optics_Together} for an illustration of this device). As measured from the common semi-minor axes, the mirrors of the vertically focusing device extend from $x=\SI{-0.5}{\meter}$ to $x=0$, and the horizontally focusing mirrors range from $x=0$ to $x=\SI{0.5}{\meter}$ along the optical axis, making the double-planar NMO $\SI{1}{\meter}$ long. \par
The combination of the desired divergence and the chosen focal length defines the required outermost semi-minor axis of the vertically and horizontally focusing NMO as follows $b_\text{0, requ.} = f \tan\left( \theta_\text{max}\right) \approx \SI{0.26}{\meter}$, yielding a beam cross-section of approximately \SI{0.52}{\meter} incident on the NMO. To ensure that each neutron can interact with at least one mirror, also for possibly larger sources, the outermost semi-minor axes of the NMOs were chosen to be $b_0 = \SI{0.3}{\meter}$.\par
A cross-shaped beamstop upstream of the double-planar NMO restricts the direct line of sight between the virtual source and the sample position. Both orthogonal segments possess a width of \SI{1.25}{\centi\meter}.
Placing an additional aperture (not shown) directly upstream of the double-planar device that restricts the illumination to a specific set of central mirrors of the NMO enables tailoring the divergence of the beam incident on the sample. \par
The NMO can only transport the desired phase space volume from the virtual source to the sample position if the source is sufficiently illuminated with the required divergence and flux. This is achieved through a parabolic feeder that focuses the low-divergence beam from the source to the maximum size of the virtual source. With an opening of $4 \times \SI{4}{\centi\meter^2}$, the upstream opening of the feeder matches the square cross-section of the guide extracting neutrons from the source. The nominal focal point of the parabola is located $\SI{4}{\meter}$ downstream from this opening with the mirrors extending $\SI{3.75}{\meter}$ along the optical axis, which results in the exit measuring $1\times\SI{1}{\centi\meter^2}$. Accounting for its spatial requirements, the centre of the pulse-cutting chopper assembly coincides with this focal point. The phase space preserving transport provided by the chosen NMO setup ensures that the beam characteristics at the virtual source are preserved and match the requirements at the sample position, i.e., a spatial extent of $10\times\SI{10}{\milli\meter^2}$ and a divergence of $2^\circ \times 2^\circ$. 
Whereas the small beam size at the sample position enables an excellent energy resolution of the secondary spectrometer, at the position of the pulse cutting chopper, the small beam size facilitates the shaping of very short neutron pulses by disc choppers, allowing the primary spectrometer to precisely match the resolution of the secondary spectrometer.

As there is not yet a defined beam port or site for the instrument, we have included an exemplary transport section connecting the reactor source and the instrument. The beam is transported by a long $m=2$ guide with a constant square cross-section of $40 \times\SI{40}{\milli\meter^2}$ and a length of $\SI{50}{\meter}$. This guide is filled up to its acceptance, corresponding to $0.5^\circ \times 0.5^\circ$ collimation at $\lambda = 2.5$~\AA, which we consider the shortest wavelength for which we optimize this cold neutron spectrometer.
This section can also be used to curve the beam and prevent the direct line of sight towards the reactor to realize a very low background level.\par
\begin{figure}[h!]
    \centering
    \includegraphics[width=\linewidth]{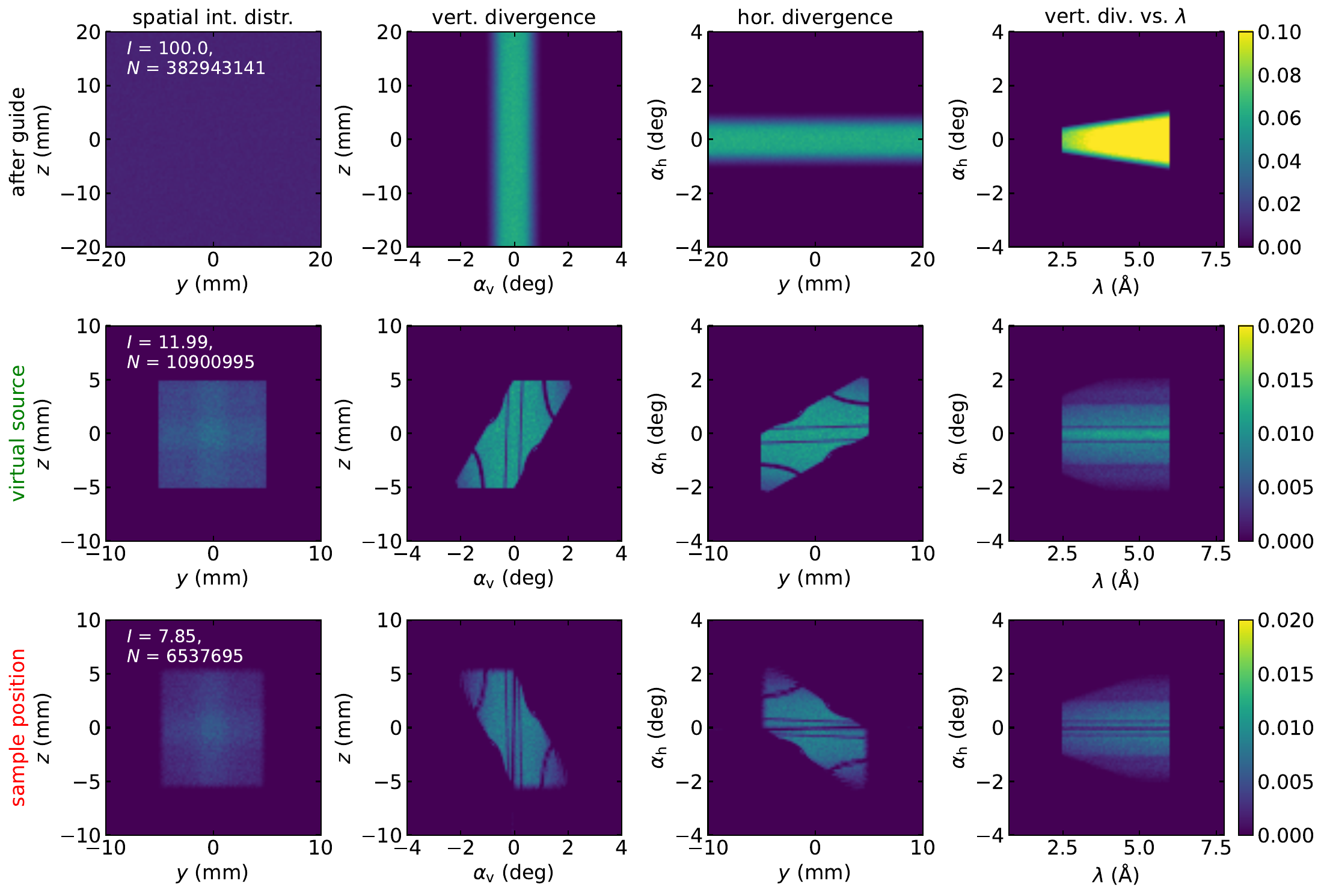}
    \caption{Simulated neutron distributions at various points along the instrument. The columns from left to right show spatial intensity distributions, neutron intensity distributions as a function of position and divergence along the vertical and horizontal directions, and the intensity as a function of the wavelength and the horizontal divergence, respectively. From top to bottom, the rows depict the distributions at the end of the long guide, at the virtual source after the pulse-cutting chopper illuminated by the parabolic trumpet, and at the sample position after the double planar elliptic NMO. The underlying coordinate system is illustrated in \cref{Fig:Mushroom_Optics_Together}. 
    }
    \label{fig:neutron_transport1}
\end{figure}

\Cref{fig:neutron_transport1} illustrates the simulated characteristics of a neutron beam with a uniform wavelength distribution ($\SI{2.5}{\angstrom} \leq \lambda \leq \SI{6.0}{\angstrom}$) at various positions along the instrument setup.
In the first row, we illustrate the spatial, angular and wavelength distributions at the end of the straight guide section.
The beam collimation aligns with the expectations for the used $m=2$ supermirror coating exhibiting a linear increase of the divergence as a function of the wavelength $\lambda$.

The second row shows analogous distributions at the position of the virtual source after passing the pulse cutting chopper with an extent of $1\times\SI{1}{\centi\meter^2}$.
The position distribution shows an enhanced, homogeneous, square-shaped intensity profile measuring $1\times \SI{1}{\centi\meter^2}$ with a slight reduction in intensity toward its corners. 
Due to the necessary minimum distance of \SI{0.25}{\milli\meter} between the exit of the parabolic guide, which is filled up to its acceptance, and the position of the pulse-cutting chopper, we observe a parallelogram-shaped volume of phase space in the horizontal and vertical direction. This phase space comprises three distinct contributions. Firstly, neutrons with minimal divergence that pass the parabolic feeder without reflection contribute to a thin rectangular phase space volume. Secondly, a diamond-shaped contribution at slightly larger divergences is attributed to neutrons reflected a single time in the feeder. Additionally, disconnected intensity islands at large divergences, distant from the optical axis, are attributed to doubly reflected neutrons.
Overall, the volume of phase space emitted at the virtual source, while not ideally rectangular, is sufficiently homogeneous within the region of interest for the instrument.

In the third row, we show the distributions at the sample position.
In particular, the spatial distribution at the virtual source and sample possess a similar shape up to a maximum divergence, validating the control of the beam size far upstream, where the products of the neutron absorption do not increase the background level in the sample and detector area.
The single reflection in the vertical and horizontal direction entails mirrored phase space volumes.
The total intensity at the sample position is reduced to 65\% compared to the intensity at the virtual source. A significant portion of the loss is attributed to the strongly divergent neutrons fed into the virtual source by multiple reflections in the parabolic feeder that  the NMO cannot transport in a single reflection. Other loss sources are the finite reflectivity of the supermirror coatings, the cross-shaped beam stop, removing very low-divergence neutrons from the beam, and the geometric losses associated with NMOs \cite{herbchristoph2022}.
Assuming a supermirror waviness of $\sigma = 0.5 \times 10^{-4}$ rad, the expected FWHM at the sample position is estimated to $10^{-4} \cdot 2.335 \cdot \SI{7.5}{\meter} \leq \SI{1.8}{\milli\meter}$, which is small relative to the sample dimensions. Other factors, such as machining imperfections and misalignment of the NMO, are more challenging to quantify and will be addressed in future experiments.


It is important to note that the design proposed upstream of the pulse cutting chopper has not been optimized towards maximizing the signal-to-noise ratio, as both the neutron source and the configuration of the neutron guide remain undetermined at this stage. The parallelogram shape of the phase space volume at the virtual source is primarily illustrative, demonstrating the phase space-preserving characteristics of the elliptic NMO. 

During the optimization process, replacing the parabolic feeder with an alternative compressing guide will enhance the instrument's performance. One promising alternative configuration under consideration facilitates a form-invariant volume transformation based on the work by Stüßer and Hofmann \cite{STUER201384}. However, the minimum distance between the parabolic feeder and the pulse-cutting chopper will still result in a parallelogram-shaped phase space volume at the position of the virtual source.
\section{Chopper system}
\label{sec:choppersystem}
\begin{figure}
    \centering
    \includegraphics[clip, width = 0.8\textwidth]{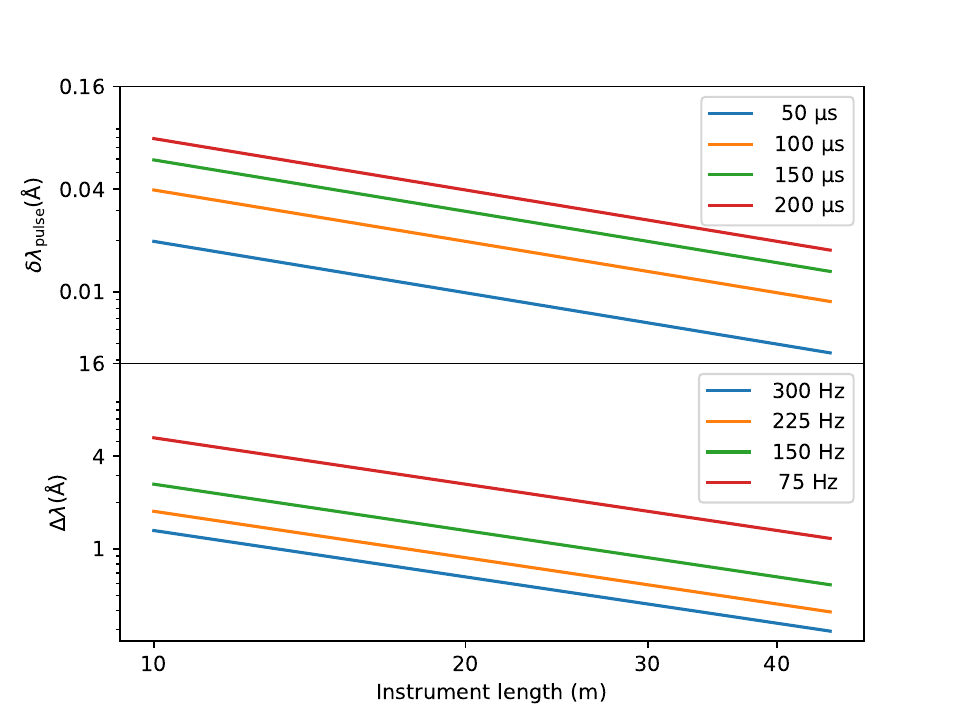}
    \caption{Primary wavelength resolution and bandwidth as a function of the instrument length for various pulse length $\tau_\mathrm{P}$ and repetition rate $f_\mathrm{RR}$. 
    }
    \label{fig:resolution_bandwidth}
\end{figure}
At a continuous neutron source, such as the FRM II, a time-of-flight crystal analyzer spectrometer can control the pulse length, the timeframe and accordingly bandwidth by means of mechanical choppers labelled  pulse shaping chopper (P), timeframe chopper (TF) and bandwidth chopper (BW).
For a given instrument length $L$, the pulse length of the P-chopper assembly $\tau_\mathrm{P}$ defines the primary wavelength resolution $\delta \lambda$ and in combination with the periodicity $T$ of the chopper system, the dynamical range, and the primary neutron bandwidth $\Delta \lambda$:
\begin{eqnarray}
\label{eq:resolution_bandwidth}
    \delta \lambda&=&\frac{h}{m_n}\frac{\tau_\mathrm{P}}{L}\\
    \Delta \lambda&=&\frac{h}{m_n}\frac{T - \tau_\mathrm{P}}{L}\approx \frac{h}{m_n}\frac{1}{f_\mathrm{RR} L} .
\end{eqnarray}
The periodicity $T$ of the chopper system is the smallest common multiple of the individual chopper periodicities.
In Fig. \ref{fig:resolution_bandwidth}, we present typical combinations of resolution and bandwidth that can be realized with state-of-the-art chopper technology  as a function of the instrument length.
Choosing an instrument length $L=15$~m we can vary the primary resolution between 0.012~\AA\, and 0.06~\AA\, i.e. close to the elastic line of the PG (002) analyzer. The primary energy resolution $\delta E/E$ can be varied between 0.5 and 2.5 \%.
At the same time, the bandwidth can be adjusted between 0.9 and 4~\AA.
The low bandwidth is perfectly suited for investigations that focus a narrow spectral range of interest with high intensity, while the large bandwidth allows overview measurements in a wide spectral range making use of the most intense part of the cold neutron spectrum.
In contrast, instruments at a pulsed neutron source are bound to a bandwidth, that is given by instrument length and the source frequency.
For the parameters of the Mushroom instrument to be built at the ISIS target station 2, this yields a huge bandwidth of approximately 14 \AA. However, depending on the required dynamic range, not the whole timeframe may contain useful information. Also the statistics will vary strongly over such a wide band.
Hence, a 'mushroom'-type instrument at a reactor makes up for the lower peak flux of the source by a more efficient use of the detector, always probing the dynamic range of interest for the specific experiment.
As a special feature, the proposed chopper system decouples the pulse length and the repetition rate completely.

\begin{figure}[h!]
    \centering
    \includegraphics[clip, width=0.48\textwidth]{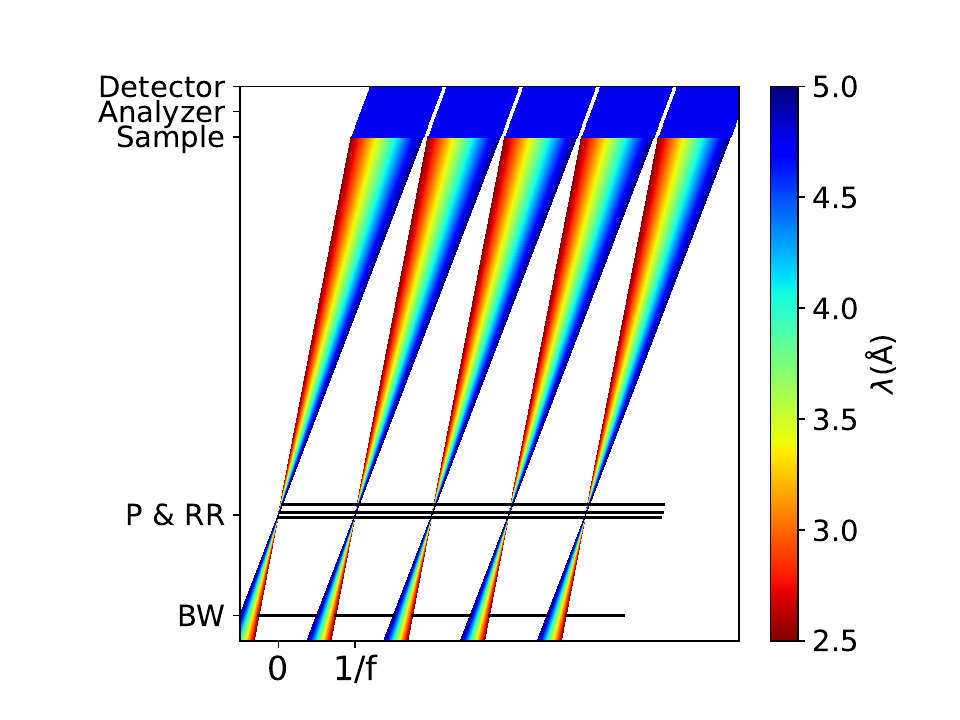}
    \includegraphics[clip, width=0.48\textwidth]{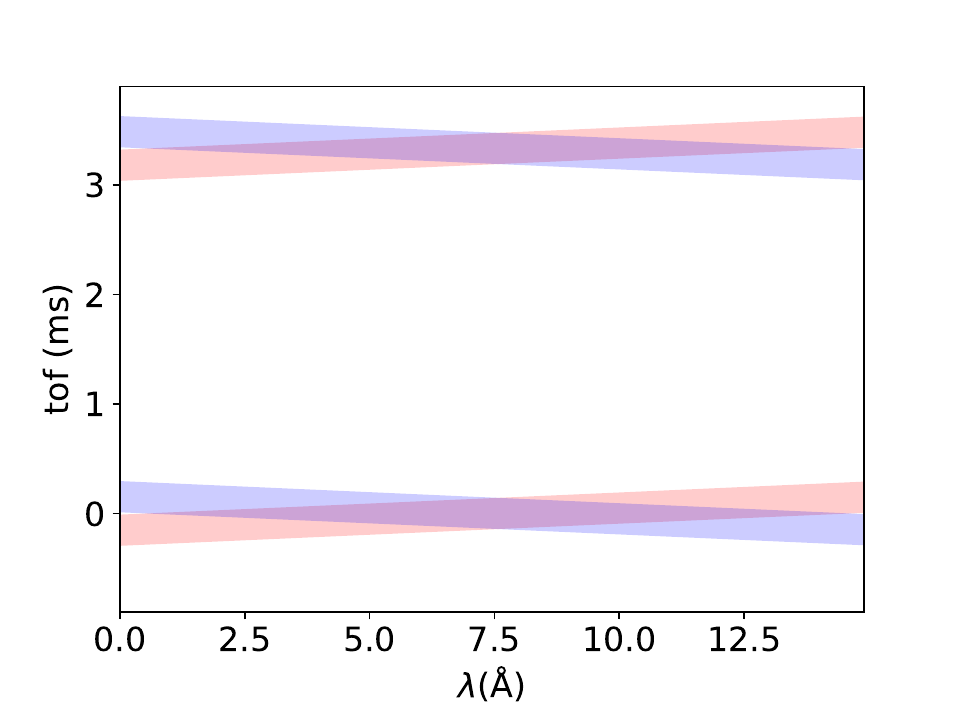}
    \includegraphics[clip, width=0.48\textwidth]{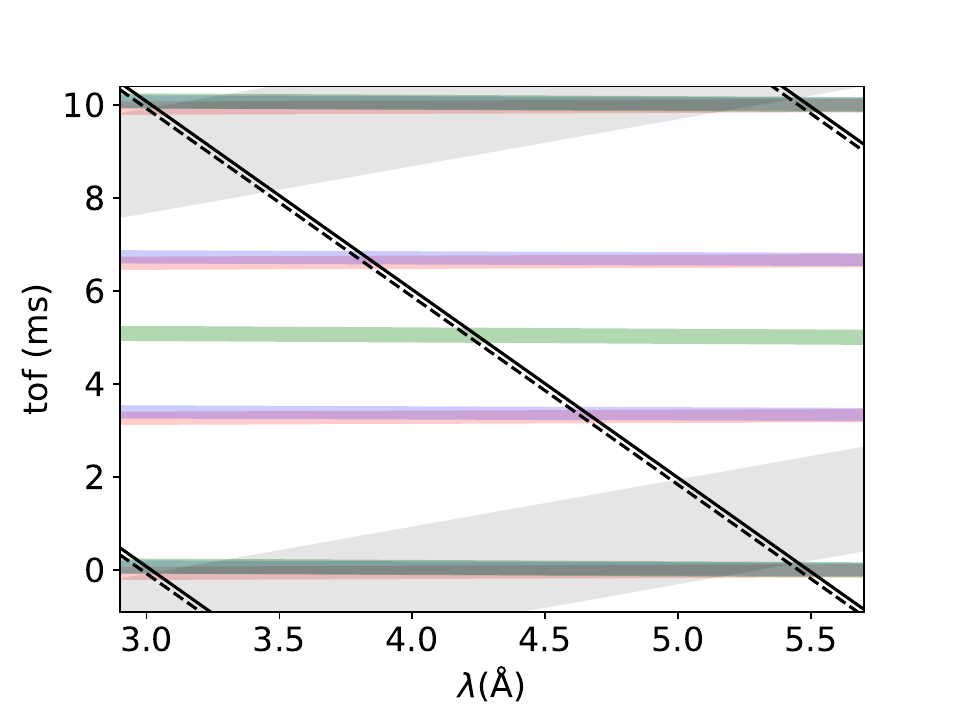}
    \includegraphics[clip, width=0.48\textwidth]{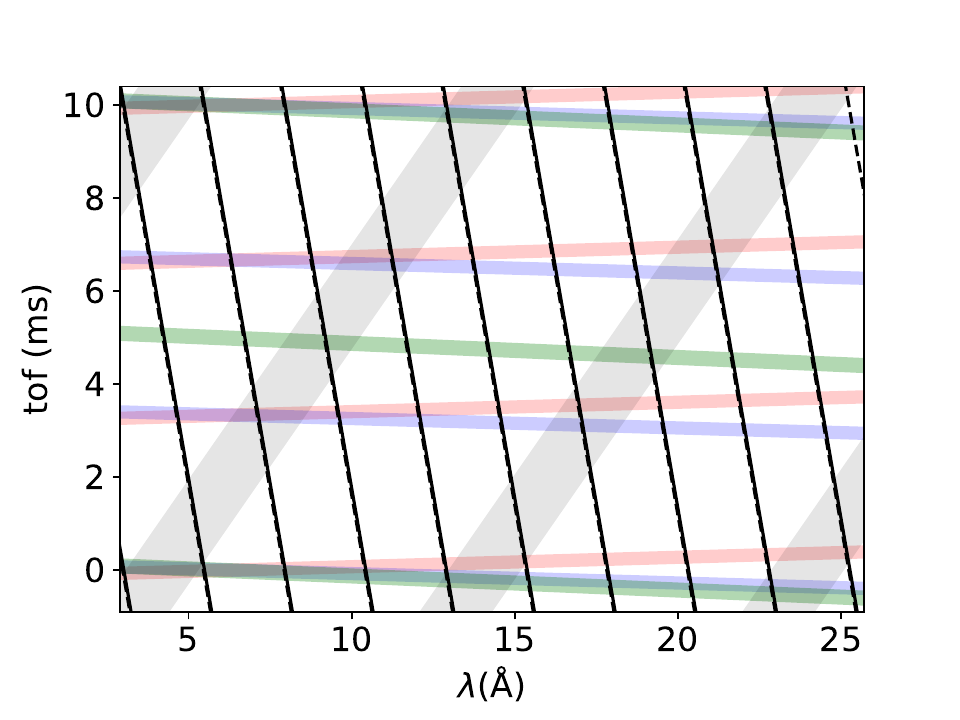}
    \caption{Top left: Flight path diagram of the instrument. Colors indicate the wavelength of the initial and analyzed neutrons, respectively.Top right: Acceptance diagram with respect to the center of the pulse shaping chopper group consisting of 2 discs spinning at $f_\mathrm{P}=300$~Hz shown in light blue and red. Bottom left: Acceptance diagram of the full chopper system covering one timeframe and the first wavelength band. Bottom right: Acceptance diagram of the full chopper system covering the wavelength range $\lambda < 26$~\AA. Acceptance of the bandwidth chopper is shown in grey, The chosen frequency combination is $f_\mathrm{P} = 300$~Hz, $f_\mathrm{RR} = 100$~Hz, $f_\mathrm{TF} = 200$~Hz.}
    \label{fig:chopper_acceptance}
\end{figure}

The implementation of short neutron pulses benefits strongly from the small beam cross-section the NMO provides at the position of the P-chopper.
For the beam of 10 mm$\times$ 10 mm, we can prepare the pulse length of 50 µs and above by two co-rotating chopper discs yielding a wavelength-dependent pulse length \cite{A.A.:1992lr, 1742-6596-746-1-012018}.
This is indicated in the time-wavelength acceptance diagram in the top left panel of Fig. \ref{fig:chopper_acceptance}, which have been introduced by Copley \cite{Copley:2003qy}.
We use the centre plane between the two P-chopper discs as the reference for the acceptance diagrams.
Also the virtual source of the NMO, which is imaged onto the sample positions, lies in this plane.
The coloured areas indicate at what time and wavelength neutrons cross the reference plane to be accepted by a chopper, i.e. they reach the chopper when the transmission is $> 0$.
We chose time zero, when the beam is open for the chosen lowest wavelength $\lambda_\mathrm{min}$ of the requested band.
The slope of the acceptance areas is proportional to the distance between the disc and the reference plane:
\begin{equation}
    \label{eq:slop_acceptance}
    m = -\frac{m_n}{h}\times S.
\end{equation}
Here $S$ indicates the position with respect to the plane containing the virtual source, which we consider also as the zero for the time-of-flight.
Choppers upstream the reference plane feature a positive slope, and the choppers downstream accordingly a negative slope.
As the two discs of the P-chopper group have the same distance from the reference plane, the absolute value $\abs{m}$ is the same.
Neutrons can pass the chopper system only when they are accepted by all choppers of the system or when all coloured regions overlap.
By controlling the distance between the choppers, one can adjust how the pulse length varies with wavelength.
In the top right panel of Fig. \ref{fig:chopper_acceptance}, we phase the 2 P-chopper discs at a distance $L_\mathrm{P_1P_2} = 0.16$~m such that the downstream disc of the group opens the beam for neutrons with wavelength $\lambda_\mathrm{min} = 2$~\AA, while the upstream chopper terminates the pulse for this wavelength.
For slightly larger wavelengths, the pulse length increases.
The maximum pulse length that can be realized is given by the width of the chopper window and the beam cross-section.
With the phase chosen in Fig. \ref{fig:chopper_acceptance}, it is achieved at $\lambda \approx 6.5 \mathrm{\AA}$. Above this wavelength, the pulse length decreases with increasing wavelength.
However, we can choose the phases of the choppers to realize any pulse length for a given neutron wavelength within this boundary.
While we can control the pulse length of the P-chopper group via the phase of the choppers, the transmission is optimized when we spin them at the highest possible speed.
We consider here a maximum frequency $f_\mathrm{max} = 300 $~Hz, which can be safely realized using carbon fibre discs and magnetically driven choppers \cite{ANDERSEN2020163402}.

We also make use of the acceptance diagrams to determine the usable bandwidth of the instrument for a periodicity of the entire chopper system.
This can be clarified from the bottom left panel of Fig. \ref{fig:chopper_acceptance}, which shows one period and one wavelength band.
To increase the periodicity of the entire system from 3.333~ms to 10~ms, the timeframe chopper (TF) runs at a frequency of 200 Hz, while the P chopper frequency is 300 Hz.  
In black, we draw here lines, which indicate the acceptance of the analyzer surface at a distance $L_\mathrm{Det}=16$~m from the reference plane.
We must be able to uniquely assign the time and wavelength of a neutron that has passed the reference plane and arrives at the analyzer.
Hence, we chose a lower wavelength $\lambda_\mathrm{min}$, which passes the reference plane at the end of the pulse, indicated by a black line.
To assign this wavelength uniquely, the analyzer should stop accepting neutrons before the same wavelength from the next pulse arrives, indicated by the dashed black line.
Therefore we reduced the periodicity by the pulse length when we calculated the bandwidth in eq. \ref{eq:resolution_bandwidth}.
This acceptance line crosses the end of the original pulse at a wavelength $\lambda_\mathrm{max}= \lambda_\mathrm{min} + \Delta \lambda$, representing the bandwidth of the instrument graphically.
To limit the band between these bounds we introduce the bandwidth chopper (BW), which has the periodicity of the P- and the TF-chopper system, e.g. 10 ms in the example in Fig. \ref{fig:chopper_acceptance}.
Its acceptance is shown in the bottom panels of Fig. \ref{fig:chopper_acceptance} as grey regions.
Its window is chosen such that it transmits only neutrons from the original pulse between the full and dashed black line as described above.
It is defined by the position of the chopper with respect to the virtual source and the analyzer and the beam cross section at the position of the BW-chopper.
For the present design we chose a distance $L_\mathrm{BW}= \frac{L_\mathrm{Det}}{4}= 4$~m.
Please note that we placed the BW-chopper upstream of the virtual source. Usually, the chopper is placed downstream of the virtual source.
Since the NMO result in a large cross-section at the position of the chopper, we shifted it symmetrically to the other side of the virtual source, where the cross-section of the feeding optics could be limited to 40~mm~$\times$~40~mm.

In the bottom right panel of Fig. \ref{fig:chopper_acceptance} we inspect the transmission of the chopper system in a wider wavelength and  time range to exclude contamination. Within this range, common acceptance regions of all choppers, i.e. colour regions of all choppers overlap, exist only for $\lambda_\mathrm{min}< \lambda < \lambda_\mathrm{max}$. In particular, the two discs of the P-chopper group will share acceptance regions only for $\lambda > 80$~\AA\, thanks to the short distance between the choppers.
To achieve a clean spectrum and, in particular, prevent frame overlap from the subsequent pulse of the bandwidth chopper, we fine-tune the distance between the second disc of the P-chopper assembly and the TF chopper to 0.12~m.
For a practical realisation of this chopper system, we would place the P-chopper discs and the TF chopper disc in a common vacuum to vary the distances between the first and the third disc up to 30~cm.

\begin{figure}
    \centering
    \includegraphics[clip, width = 0.49\textwidth]{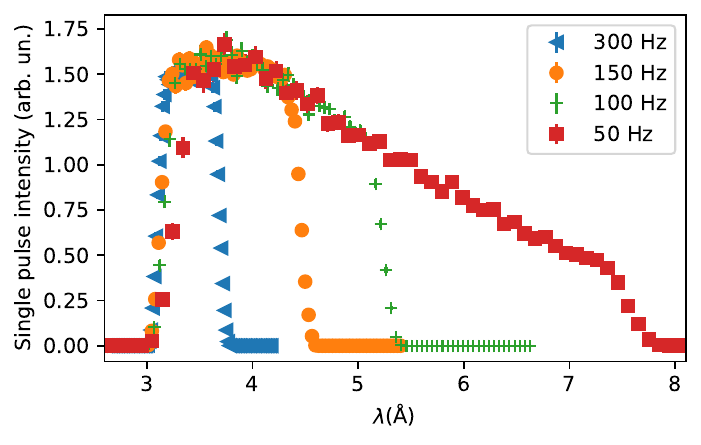}
    \includegraphics[clip, width = 0.49\textwidth]{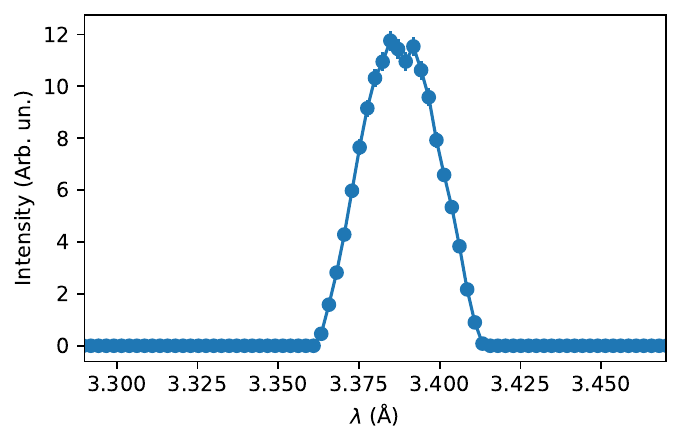}
    \caption{Left: Single pulse wavelength spectrum at the sample position for different frequencies of the BW-chopper. 
    Right: Wavelength distribution within a 40 $\mu$s time window. The P-chopper frequency is 300 Hz, the TF frequency is 300 Hz, 150 Hz, 100 Hz and 50 Hz, respectively.
}
    \label{fig:sim_bandwidth}
\end{figure}

The presented chopper system provides continuous control of the pulse length via the phases of the P-chopper assembly, but also a broad range of bandwidth and, hence, a flexible dynamic range.
Changing the bandwidth hardly affects the sample flux, as can be seen from the left panel in Fig. \ref{fig:sim_bandwidth}. 
It shows the spectrum of a single chopper pulse at the sample position. Within the statistical limits of the simulation, the spectral intensity is the same for most of the bandwidth.
The sample flux is given by the integration of the spectrum and then multiplication with the frequency.
For a large frequency, the intensity is concentrated within a narrow dynamical range, while a large dynamical range is accessible for low frequency.
For higher repetition rate, the edges of the transmitted spectrum are steeper, as the bandwidth choppers spin faster.
The wavelength resolution at the sample position is presented in the right panel of Fig. \ref{fig:sim_bandwidth} as calculated from the 2-dimensional wavelength-time-of-flight distribution, integrating over a time window of 40 $\mu$s. 
The trapezoidal pulses shaped by the P choppers result in a symmetric wavelength distribution that does not exhibit any tails. This gives rise to a clean resolution function. 

\section{Secondary crystal spectrometer}\label{sec:secondary_spectrometer}
\subsection{Working principle}
\begin{figure}[ht]
	\centering
	\includegraphics[width=0.5\textwidth]{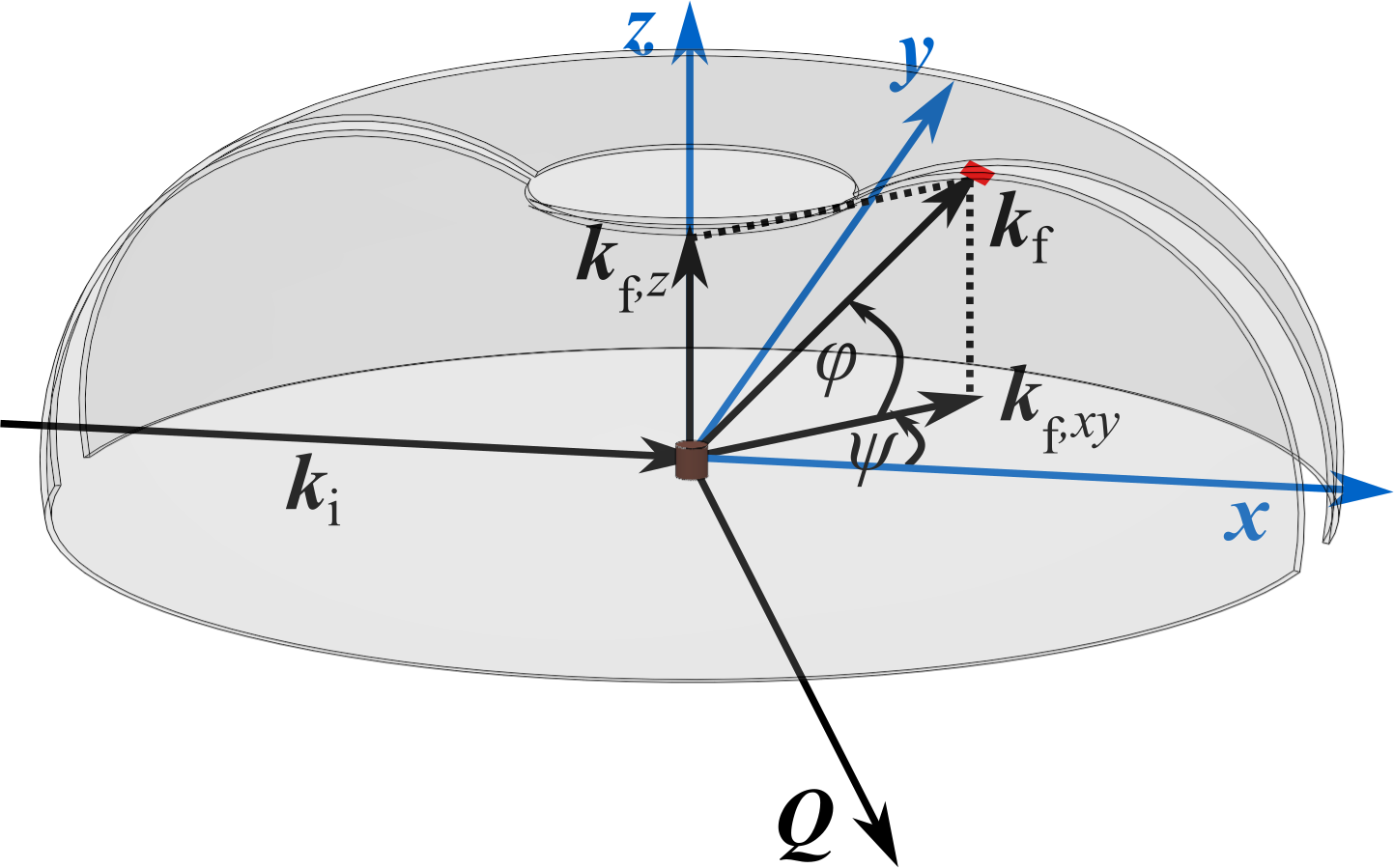}
	\caption{A schematic of the analyser illustrating the coordinate systems. The Cartesian system is defined with its origin at the sample position, the $z$-axis pointing upwards, the $x$-axis being parallel to the wavevector $\bm{k}_{\mathrm{i}}$ of the incoming beam, and the $y$-axis completing a right-hand system. The analyser has a mirror symmetry with regard to the $xz$-and $yz$-planes, with the angular position $(\psi,\varphi)$-of one of its segments highlighted in red. For simplicity, the analyser is depicted as a singular entity on either side of the direct beam, adopting an elliptical form in a vertical sectional view. Consequently, the azimuthal angle $\psi$ signifies the angle between the wavevector $\bm{k}_{\mathrm{f}}$ of scattered beam and the $xz$-plane, while the polar angle $\varphi$ denotes the angle between $\bm{k}_{\mathrm{f}}$ and the $xy$-plane.}
	\label{Fig:Mushroom_Analyser}
\end{figure}

The secondary spectrometer with a mushroom-like super-flatcone analyser is the unique feature of the Mushroom concept. The analyser consists of a large number of segments made of pyrolytic graphite (PG), reflecting the neutrons according to Bragg's law after their interaction with the sample. The position of a segment is denoted by the azimuthal (horizontal) $\psi$ and polar (vertical) $ \varphi$ angles of its centre with respect to the sample position. The angles and the wavenumbers in the coordinate system are illustrated in Fig. \ref{Fig:Mushroom_Analyser}, while other components of the secondary spectrometer are shown in Fig. \ref{Fig:Mushroom_Optics_Together}. A measurement with the sample environment and a vanadium sample can help to subtract the background signals due to factors such as spurious scattering by the cryostat or inelastic scattering by the PG analyser. A radial collimator around the sample will also reduce this background.

\begin{figure}[ht]
	\centering
	\includegraphics[width=0.8\textwidth]{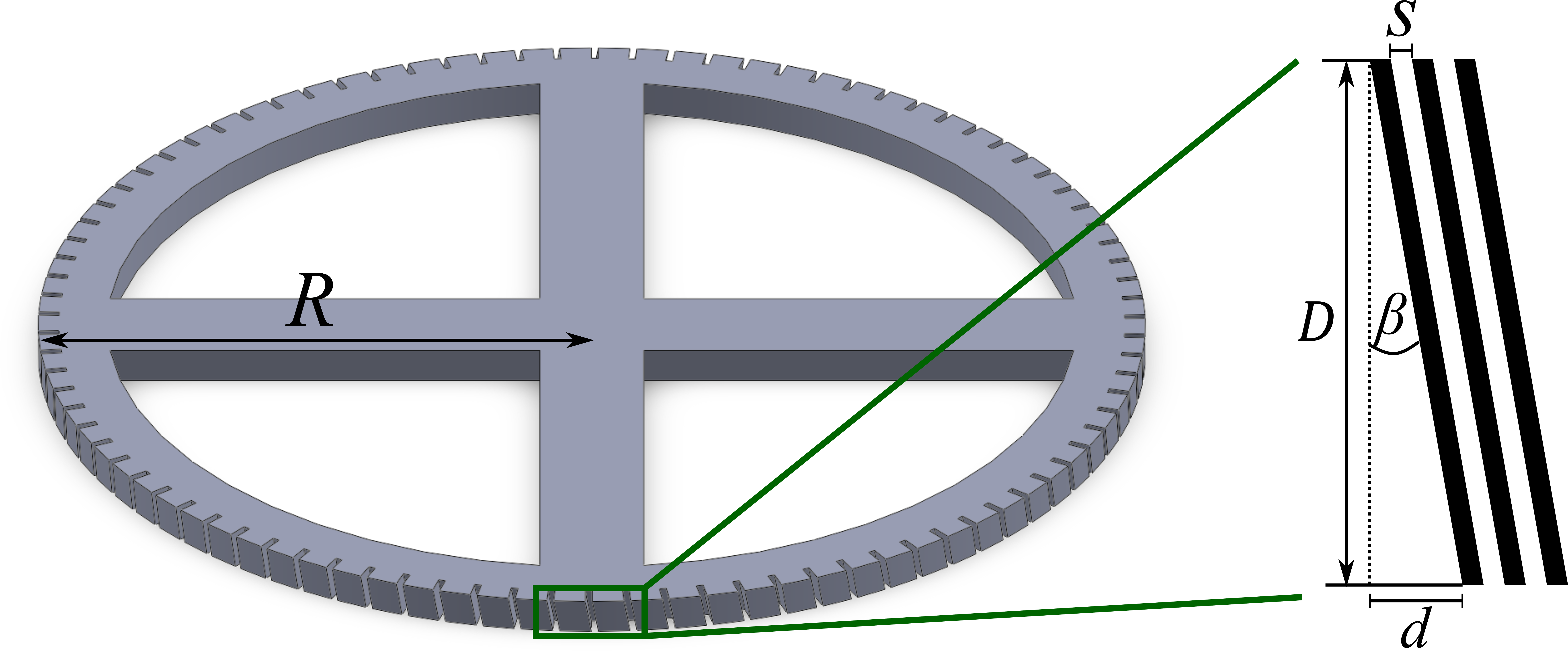}
	\caption{A simplified schematic of the order selector, featuring a radius $R$ and thickness $D$, for the secondary spectrometer of Mushroom. The size and number of slits are adjusted for the schematic to improve visibility, with the enlarged version illustrated on the right-hand side. The geometry of a slit is defined by its width $s$, tilting angle $\beta$, and phase difference $d=D\tan{\beta}$ between its two ends.}
	\label{Fig:Mushroom_OrderSelector}
\end{figure}

Since the reflection by the analyser has the contributions mainly from the diffraction orders PG 002 and 004, an order selector is implemented to select the diffraction order $n_{A}$ for determining the wavenumber $k_{\mathrm{f}}$ unambiguously. This is realised by using a mechanical velocity selector with a slit at the focal point of an elliptic curve in a vertical view, with the other focal point located at the sample position. The order selector is depicted in Fig. \ref{Fig:Mushroom_OrderSelector}, with a radius $R$ and a thickness $D$. Each slit has a width $s$, an off-vertical angle $\beta$, and a phase distance $d=D \tan{\beta}$, which means that the time of the plate in the distance $d$ corresponds to the time of the neutron beam in the distance $D$. Denoting the rotation frequency $f$, the parameters are connected via $\frac{D}{v_{0}} = \frac{d}{2\pi Rf}$ or $\tan{\beta} = \frac{2\pi Rf}{v_{0}}$. The relative resolution can be expressed as
\begin{equation}
	u_{\mathrm{r}} = \frac{s}{d}.    
\end{equation}
For distinguishing between the orders PG 002 and PG 004, a resolution of $u_{r}\approx\SI{25}{\percent}$ is already satisfactory.

\begin{figure}[ht]
	\begin{subfigure}{0.4\textwidth}
		\centering
		\includegraphics[height=0.27\textheight]{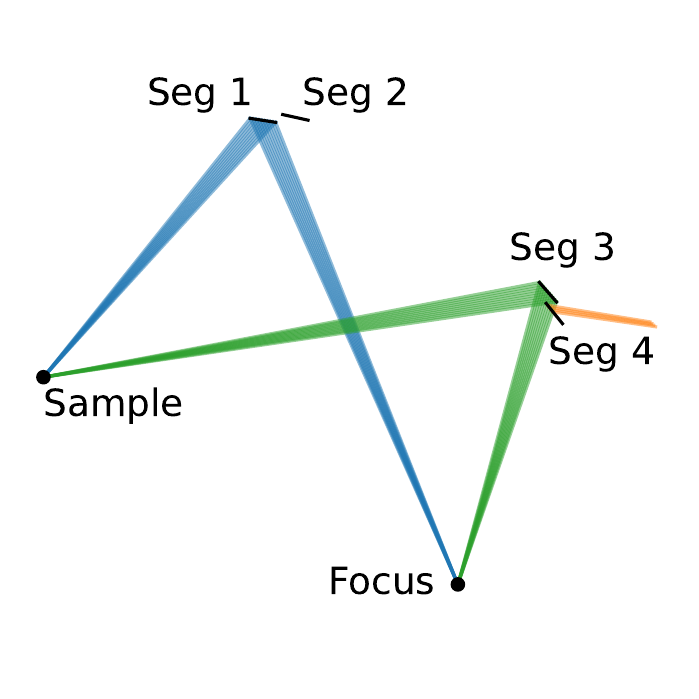}
		\caption{}
		\label{Fig:Mushroom_Analysers_DoubleReflections}
	\end{subfigure}
	\begin{subfigure}{0.6\textwidth}
		\centering
		\includegraphics[height=0.27\textheight]{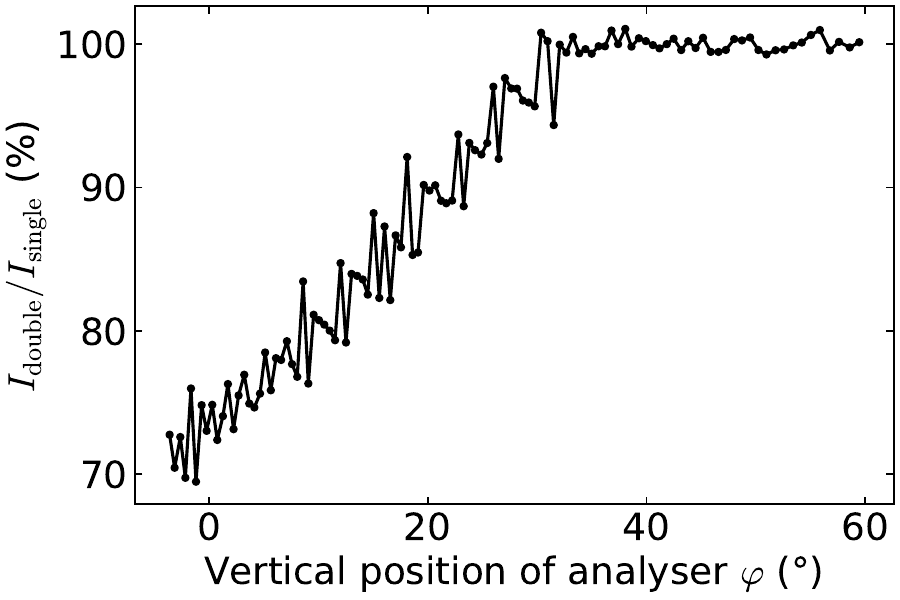}
		\caption{}
		\label{Fig:Mushroom_Intensities_Doubles}
	\end{subfigure}
	\caption{Panel a: Vertical view visualising possible multi-reflections due to neighbouring analyser segments. Neutrons reflected by the 1st segment (``Seg 1'' are not reflected by the 2nd segment (``Seg 2''), as shown by the blue trajectory. Some neutrons reflected by the 3rd segment (``Seg 3'') may be reflected by the 4th segment (``Seg 4''), depicted by orange and green for beams with and without double reflections, respectively. Panel b: Simulations comparing the intensity of neutrons reflected by a single analyser segment ($I_{\mathrm{single}}$) with the intensity where the neutron beam reflected by this segment may also be reflected by the next one ($I_{\mathrm{double}}$). Further details on the simulations are provided in the text.}
	\label{Fig:Mushroom_Analysers_DoubleSegments}
\end{figure}

The segments are aligned in a Rowland geometry, where the centres of all analyser segments (A) are on the same circle as the sample position (S) and another focal point (F). The normal of each segment bisects the angle $\angle\mathrm{SAF}$, providing the same scattering angle at the centres of all segments. This design allows measuring a constant wavenumber $k_{\mathrm{f}}$, while $k_{\mathrm{f},z}$, the $z$-component of the wavevector $\Vec{k}_{\mathrm{f}}$, is determined by the vertical angle $\varphi$. A disadvantage of this design is the shadowing effect when the segments are closer to the velocity selector than to the sample position. Here, a segment can be in the way of the neutrons reflected by the previous segment. Hence, some neutrons can be reflected again by the lower segment. This effect is simulated and depicted in Fig. \ref{Fig:Mushroom_Analysers_DoubleReflections}, where the intensity considering two neighbouring segments is compared to the situation with only one segment. The simulation shows is no intensity loss at high vertical positions due to the absence of shadowing, while the intensity reduces down to approximately \SI{70}{\percent} at the lowest segment. This effect can be tolerated but the intensity has to be corrected in a measurement. 

\begin{figure}[ht]
	\centering
	\includegraphics[width=0.6\textwidth]{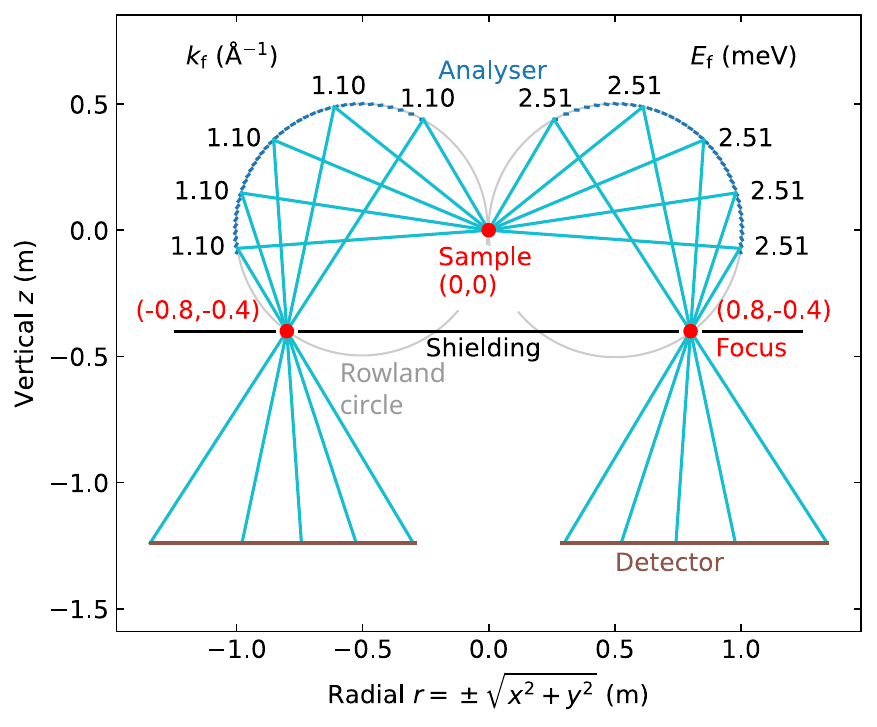}
	\caption{Schematic of the Mushroom secondary spectrometer in a sectional view. The dark blue dashed lines denote the analyzer segments aligned on a Rowland circle, grey (due to a large number of segments, only every third is shown). The light blue lines showcase some representative trajectories of the neutron beam, the brown lines illustrate the PSD banks, the black line indicates the shielding, and the red points depict the sample position at $(0, 0)$ as well as the other focal points at $(\pm\SI{0.8}{\metre},\SI{-0.4}{\metre})$. The geometrical parameters correspond to a configuration for $k_{\mathrm{f}} = \SI{1.1}{\per\angstrom}$ and $E_{\mathrm{f}} = \SI{2.51}{\milli\electronvolt}$.}
	\label{Fig:Mushroom_Geometry_2DVertical}
\end{figure}

In order to keep a compact size of the secondary spectrometer and to offer a space for a cryostat with a diameter of $\geq\SI{50}{\centi\metre}$ at the centre, a geometry corresponding to $k_{\mathrm{f},002}=\SI{1.1}{\per\angstrom}$ at PG 002 ($k_{\mathrm{f},004}=\SI{2.2}{\per\angstrom}$) is chosen. With the configuration shown in Fig. \ref{Fig:Mushroom_Geometry_2DVertical}, the analyser covers the angle $\varphi$ from \SIrange{-5}{60}{\degree}, and $\psi$ in the range of from \SIrange{-170}{-10}{\degree} and from \SIrange{10}{170}{\degree}. The sample position is denoted as the centre $(0,0)$, and the other focal point is placed at $(\SI{0.4}{\metre},\SI{-0.8}{\metre})$ in a vertical view. The PSD bank with the outer radius of \SI{1.35}{\metre} is \SI{1.24}{\metre} below the sample position. The spatial resolution of the PSD He-tubes is in the same order of magnitude as the size of analyser segments (\SI{1}{\centi\metre}). For the order selector, one can set $\gamma=\SI{12}{\degree}$, giving the frequency for PG 002 $f_{002}\approx\SI{29.3}{\hertz}$ and for PG 004 $f_{004}\approx\SI{58.6}{\hertz}$. With $s=\SI{3}{\milli\metre}$, one has $d=\SI{12}{\milli\metre}$ and $D\approx\SI{56}{\milli\metre}$.

\subsection{Resolution}
The resolution of the secondary spectrometer of Mushroom can be characterised by three factors, the uncertainty of the wavenumber, $\Delta k_{\mathrm{f}}$, the angular spread of the wavevector $\bm{k}_{\mathrm{f}}$ in the horizontal plane, $\Delta\psi_{\mathrm{f}}$, and the angular spread in the vertical plane, $\Delta\varphi_{\mathrm{f}}$. The uncertainties are characterised by the respective standard deviations (SDs).

\begin{figure}[ht]
	\centering
	\includegraphics[width=.75\linewidth]{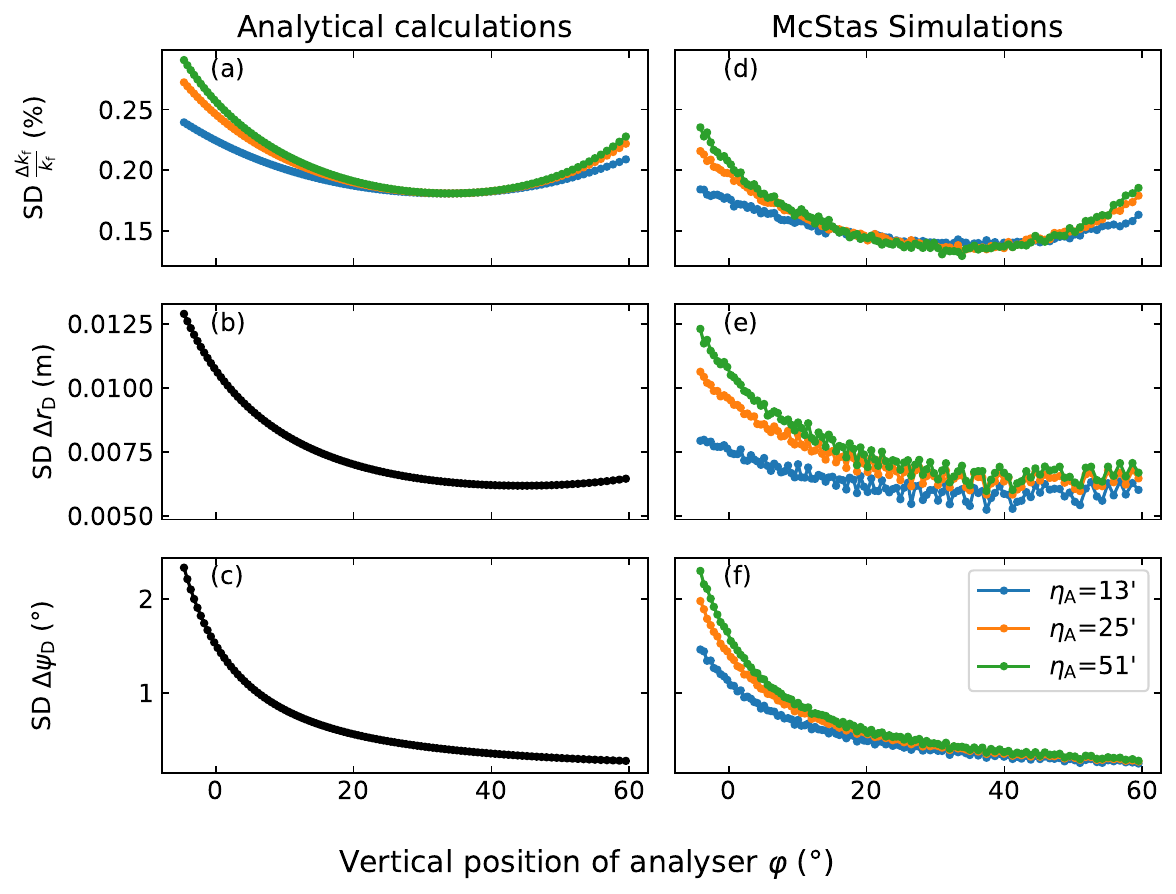}
	\caption{The $\bm{k}_{\mathrm{f}}$-resolution of the secondary spectrometer of the Mushroom, where the radial position is defined as $r=\pm\sqrt{x^2+y^2}$, with $x$ and $y$ defined in the Cartesian coordinate system in Fig. \ref{Fig:Mushroom_Analyser}. Panels a and d: standard deviation (SD) of the wavenumber $k_{\mathrm{f}}$ in percentage. Panels b and e: SD of the radial position ($r_{\mathrm{D}}$) on the PSD. Panels c and f: SD of the angular position ($\psi_{\mathrm{D}}$) on the PSD. Panels a, b and c depict analytical calculations, while panels d, e and f present simulations using McStas. The values of mosaicity $\eta_{\mathrm{A}}$ of analyser crystals are labelled in colour, with the three SD values of \SI{13}{\arcminute}, \SI{25}{\arcminute} and \SI{51}{\arcminute} corresponding to \SI{0.5}{\degree}, \SI{1}{\degree} and \SI{2}{\degree} characterised by FWHM, respectively. The black curve in $\Delta r_{\mathrm{D}}$ and $\Delta\psi_{\mathrm{D}}$ indicate that the respective analytical calculations do not involve $\eta_{\mathrm{A}}$.}
	\label{Fig:Mushroom_SD_1cm_PSD}
\end{figure}

An analytical estimation of the wavenumber resolution, $\Delta k_{\mathrm{f}}$, can be derived from Bragg's law, yielding:
\begin{equation}
	\frac{\Delta k_{\mathrm{f}}}{k_{\mathrm{f}}} = \sqrt{\left(\frac{\Delta d}{d}\right)^{2} + \left(\cot{\theta_{\mathrm{A}}}\Delta\theta_{\mathrm{A}}\right)^{2}}. \label{Eq:Mushroom_Uncertainty_kf}
\end{equation}
The $d$-spacing of a PG crystal was determined by T. Keller \textit{et al.} to exhibit a broadening of $\frac{\Delta d}{d}\lessapprox\num{2.5e-4}$ (characterised by its SD) \cite{Keller2002}. The uncertainty of the scattering angle at an analyser segment can be expressed as \cite{Kalus1973,Telling2005}:
\begin{equation}\label{Eq:Mushroom_AngularSpread_Analyser}
	\Delta\theta_{\mathrm{A}} = \sqrt{\frac{\alpha_{\mathrm{A,i}}^{2}\alpha_{\mathrm{A,f}}^{2} + \eta_{\mathrm{A}}^{2}\alpha_{\mathrm{A,i}}^{2} + \eta_{\mathrm{A}}^{2}\alpha_{\mathrm{A,f}}^{2}}{4\eta_{\mathrm{A}}^{2} + \alpha_{\mathrm{A,i}}^{2} + \alpha_{\mathrm{A,f}}^{2}}},
\end{equation}
where $\eta_{\mathrm{A}}$ denotes the mosaicity of the PG crystal used in the analyser, ranging typically from \SIrange{0.13}{1.3}{\degree} (characterised by its SD) \cite{Ohler1997}. The incoming and outgoing divergences concerning the analyser are signified by $\alpha_{\mathrm{A,i}}$ and $\alpha_{\mathrm{A,f}}$, respectively. They can be estimated by considering orientation spreads of the respective trajectory of the neutron beams due to the finite sizes of the sample $l_{\mathrm{S}}$, the analyser segment $l_{\mathrm{A}}$ and the opening of the focal point $l_{\mathrm{F}}$, yielding:
\begin{equation}
    \begin{split}
        \alpha_{\mathrm{A,i}} &= \frac{\theta_{A,\mathrm{i},\mathrm{max}}-\theta_{A,\mathrm{i},\mathrm{min}}}{\sqrt{24}} \\
        \alpha_{\mathrm{A,f}} &= \frac{\theta_{A,\mathrm{f},\mathrm{max}}-\theta_{A,\mathrm{f},\mathrm{min}}}{\sqrt{24}}.
    \end{split}
\end{equation}
Here, $\theta_{A,\mathrm{i},\mathrm{max}}$ and $\theta_{A,\mathrm{i},\mathrm{min}}$ denote the maximum and minimum incident angles at the analyser (zero mosaicity), respectively, and $\theta_{A,\mathrm{f},\mathrm{max}}$ and $\theta_{A,\mathrm{f},\mathrm{min}}$ signify the outgoing angles at the analyser. The distribution of the incident and outgoing angles at the analyser is each approximated by a symmetric triangular distribution, yielding the SD $\sigma=\sqrt{\frac{L}{24}}$, with the width of the triangle $L$. The analytical result is shown in Fig. \ref{Fig:Mushroom_SD_1cm_PSD}a, considering three values of mosaicity $\eta_{\mathrm{A}}$ as well as $l_{\mathrm{S}}=l_{\mathrm{A}}=l_{\mathrm{F}}=\SI{1}{\centi\metre}$.

The angular spreads $\Delta\psi_{\mathrm{f}}$ (horizontal) and $\Delta\varphi_{\mathrm{f}}$ (vertical) of $\bm{k}_{\mathrm{f}}$ can be derived from the detector spreads with beam divergence. The detector spreads in the radial ($\Delta r_{\mathrm{D}}$) and azimuthal ($\Delta\psi_{\mathrm{D}}$) dimensions are depicted in Fig. \ref{Fig:Mushroom_SD_1cm_PSD}b and \ref{Fig:Mushroom_SD_1cm_PSD}c, where the maximum divergences between the analyser and velocity selector permitted by the geometry are considered, while the mosaicity of analyser crystals is not involved in the calcualtion.

In addition to analytical calculations, simulations were carried out using McStas for $l_{\mathrm{S}}=l_{\mathrm{A}}=l_{\mathrm{F}}=\SI{1}{\centi\metre}$, depicted in Fig. \ref{Fig:Mushroom_SD_1cm_PSD}. Here, the SD of the wavenumber $k_{\mathrm{f}}$ in percentage ($\frac{\Delta k_{\mathrm{f}}}{k_{\mathrm{f}}}$), as well as the radial ($\Delta r_{\mathrm{D}}$) and angular ($\Delta\psi_{\mathrm{D}}$) spreads on the PSD are compared.

Fig. \ref{Fig:Mushroom_SD_1cm_PSD} shows a good agreement between the analytical calculations and simulations qualitatively. The panels a and d indicate a very good wavenumber resolution of the secondary spectrometer, giving a relative uncertainty of about \SI{0.2}{\percent}. This is the case even with a relaxed mosaicity (FWHM) $\eta_{\mathrm{A}}\geq\SI{1}{\degree}$ of analyser crystals. The radial spread on the detector is around \SI{1}{\centi\metre}, corresponding to a vertical spread $\Delta\varphi_{\mathrm{f}}$ around \SI{0.5}{\degree}, while the horizontal spread $\Delta\psi_{\mathrm{f}}$ reaches \SI{2}{\degree} when neutrons are reflected by the analyser segments with low $\varphi$. The phenomenon at low $\varphi$-values results from the short distance between the analyser and the velocity selector, about \SI{0.4}{\metre} at the lowest segment. Besides, the neutrons coming from the analyser at a lower $\varphi$ arrive at the PSD with a smaller radial position, resulting in a larger angular spread than at a larger radial position considering the same tangential spread in length. This can be improved by implementing large crystals formed in the shape of circular stripes, which have a rotational symmetry at the same vertical position. With this, the horizontal spread $\Delta\psi_{\mathrm{f}}$ can be improved to below \SI{1}{\degree}. The mechanical possibility is under further investigation. Comparisons between Fig. \ref{Fig:Mushroom_SD_1cm_PSD}b and Fig. \ref{Fig:Mushroom_SD_1cm_PSD}e as well as between Fig. \ref{Fig:Mushroom_SD_1cm_PSD}c and Fig. \ref{Fig:Mushroom_SD_1cm_PSD}f show a decrease of detector spreads at finer mosaicities of analyser crystals. This agrees that a small mosaicity accepts a smaller part of the beam divergence as a relaxed mosaicity does.

\section{Conclusion}

Here, we have investigated the potential of a time-of-flight crystal analyser spectrometer at a continuous neutron source.
The instrument uses the large angular acceptance of PG to increase the count rate on the detector.
At the same time a good energy resolution is achieved due to the good collimation realized by the illuminated sample size, the width of the individual analyzer crystals and the position resolution of the detector.
Therefore the control of the beam spot size at the sample position is crucial for the energy and momentum resolution of the instrument.
Thanks to the NMO, this beamspot size is effectively controlled at the position of the virtual source far upstream of the sample area.
We have shown that we can effectively transport exactly the phase space element required by the instrument.
So the NMO allows to limit the illumination to the sample only, preventing any background from neutrons that are scattered by the sample environment. 
Furthermore, all equipment needed to tailor the beam properties are far upstream.
This gives excellent access to the sample area facilitating the operation of very complex sample environment, but also providing space for ancillary equipment, e.g. to control the state of the sample.
But also the primary resolution benefits strongly from the NMO.
It provides a natural position for the pulse shaping chopper.
Thanks to the small cross section a trapezoidal transmission profile can be realized by a single chopper disc.
As the pulse length requirements to match the secondary instrument resolution are well within the range, that can be realized by state-of-the-art chopper technology, we can use the concept of optical blind choppers.
So we relax the pulse length for the longer wavelength in the band and compensate such the reduction of the source brightness towards longer wavelengths.

While the use of choppers seems counnter-intuitive on a continuous source such as the FRM II, it features distinct advantages for the study of dynamics in solids.
By means of the repetititon rate we can control the bandwidth of the instrument freely.
The bandwidth can be tuned from 0.8~\AA~to 4~\AA. Tuning the lower limit of the bandwidth to the elastic line of the analyser, the instruments covers the dynamic range 0~meV~$< \hbar \omega <20$~meV in the wide band configuration.
If it is tuned symmetric to the elastic line in the narrow band configuration, the intensity is concentrated in the dynamic range -0.4~meV$< \hbar \omega <$~0.4~meV, but can also be shifted to any other energy transfer range of interest.
The bandwidth considerations are completely decoupled from the energy resolution requirements. Thanks to a narrow virtual source, the energy resolution can be controlled solely by the phase of the P choppers.
The combination of the optics and the chopper system provides a sample flux, that reaches 10$^8$~neutrons$/$cm$^2/$s. 

We are convinced, that this type of instrument with a huge acceptance of the analyser and a high sample flux in combination with a large accessible area around the sample will be useful for a wide range of applications in particular in the field of quantum and topological materials.

\section{Acknowledgement}
Fruitful discussions about the crystal secondary mushroom concept with R.~Bewley, ISIS are highly acknowledged. We also would like to thank one of the reviewers for their helpful comments on the instrument proposal. This work was funded by the BMBF (German Federal Ministry for Research) under project number ErUM-Pro/05K19WOA.
\bibliographystyle{JHEP}
\bibliography{main}
\end{document}